\documentclass{article}
\usepackage{frascatiphys}
\usepackage{graphicx}
\usepackage{hyperref}
\usepackage{amssymb}

\newcommand{\alp}{a}
\newcommand{\ga}{g_{a\gamma}}
\begin{document}
\title{ 
DARK SECTORS AT FIXED TARGETS: THE EXAMPLE OF NA62
}
\author{
Babette D\"obrich\thanks{ On behalf of the NA62 Collaboration:
R.~Aliberti, F.~Ambrosino, R.~Ammendola, B.~Angelucci, A.~Antonelli, G.~Anzivino, R.~Arcidiacono, 
M.~Barbanera, A.~Biagioni, L.~Bician
% ...
}        \\
{\em CERN, 1211 Geneva 23, Switzerland} \\
}
\maketitle
\baselineskip=11.6pt
\begin{abstract}
If new physics manifests itself in the existence of
very weakly coupled particles  of MeV-GeV mass-scale, fixed-target experiments 
can be an excellent instrument
to discover it. In these proceedings, we review
especially the sensitivity of the NA62 experiment to this physics scenario.
\end{abstract}
\baselineskip=14pt

\section{Weakly coupled particles at fixed target set-ups}

Albeit we can describe our findings about elementary particles and their interactions
to an incredible precision in the
`Standard Model of particle physics' (SM),
it is clear that the particle content therein is likely not complete.
One of the most blatant evidence for our insufficient knowledge is:
we do not know what Dark Matter (DM) particles (constituting
$\sim$80\% of
all matter) are.
Exploration of the existence of particles at high 
energies (masses) is and will be performed,
e.g., by the LHC.
However, 
new particles might also be found at much lower energy scales 
but interaction strengths which are very tiny.
If such particles exist, their feeble
coupling would make them comparably long-lived and 
thus they could escape strong constraints from searches with colliders.

To search such long-lived particles,
independently of whether they explain DM, a number existing
and proposed fixed-target/beam-dump experiments have
set up corresponding programs.
Amongst the ones that exploit
a high-energy proton beam as primary beam
are the proposed SHiP experiment,
as well as running experiments SeaQuest at Fermilab and NA62 at CERN \cite{NA62:2017rwk}. Results and 
prospects of the latter will be subject of this article,
albeit some of our plots of Sect.~\ref{sec:dump} also
include prospects for other experiments.
Comprehensive projections for Seaquest can be found in \cite{Berlin:2018pwi} and
for SHiP in \cite{Anelli:2015pba}.

In general, a primary beam of protons,
electrons or even muons is used to 
produce such long-lived particles of a `dark' or `hidden' sector,
which is motivated by different BSM physics, e.g. \cite{Essig:2013lka,Alekhin:2015byh,Alexander:2016aln}. 
Typically, a higher primary beam energy is favored
to achieve sizable production cross-sections. The experiment
geometry then systematically shapes the accessible
parameter space in the coupling-mass plane as it `selects' the longevity of the
particles that can be searched for. Importantly, the number
of primary particles correlates with the feebleness 
of particles that can be probed.

If the exotic particles are stable and thus invisible for the experiment
due to their small coupling, they could be found
by missing-mass or missing energy techniques.
Examples of such searches at NA62 are presented in sect. \ref{sec:k_decay}.
 
If the new particles decay, their final states will guide
us in pin-pointing the new responsible interaction, see examples
in sect. \ref{sec:dump}.

\section{NA62 at CERN's SPS}

\begin{figure}[]
\begin{center}
 \includegraphics[width=1\textwidth]{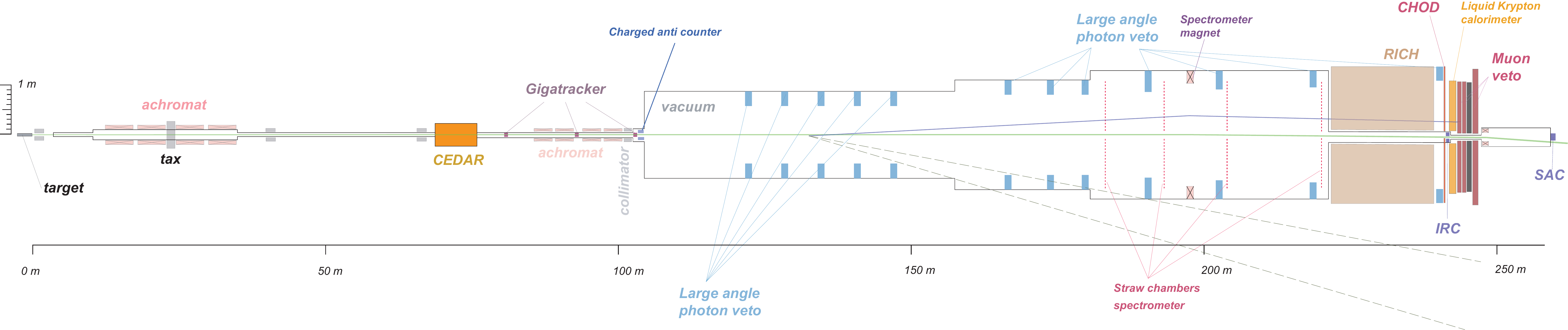}

 \caption{\label{fig:na62} \it
 Layout of NA62. 
 See \cite{NA62:2017rwk}
 for details on the detector. For closed-collimator runs the proton beam is dumped
 at $\sim$ 23m (w.r.t. to the nominal target position at 0m, relevant in Kaon-mode).}
 \end{center}
\end{figure}

The NA62 experiment aims at a precise measurement of the rare decay 
 $K^+\rightarrow \pi^+ \nu \bar{\nu}$. As the SM branching ratio
 of this decay is extremely small\cite{Buras:2015qea} $\mathcal{O}(10^{-10})$, the experiment
 is equipped with a hermetic detector system, cf. fig. \ref{fig:na62}.
In addition, the experiment achieves a $\mathcal{O}(100)$ps
 timing resolution. 
 
 The SPS primary 400 GeV proton beam interacting in
an upstream beryllium target (at 0m in fig. \ref{fig:na62})
produces a 75 GeV unseparated secondary beam (containing around
6 \% Kaons) for NA62,
selected by an achromat
around 23m downstream the target. This beam is guided through
a beamline into the experimental hall, with the first
detector (KTAG) located at around 70 m.

Two trackers: The GTK (Si-pixel), and the STRAWs
 allow to determine the 
 3-momentum of the incident particles and their decay products, respectively.
 The GTK data is matched with the KTAG (differential Cerenkov
counter) to obtain the full Kaon 4-momentum. The CHANTI station 
provides protection by vetoing inelastic 
interactions of the 75 GeV beam in the third GTK tracker-station. A RICH
positively identifies secondary charged pions.
Further Hadron ID is provided by the calorimeters MUV1 and MUV2 .
To veto unwanted decay modes,  Muon ID is provided by the MUV3 plastic
scintillator detector, placed after an iron absorber.
Finally, photons can be vetoed at small angles by the IRC and SAC,
at intermediate angles by the liquid krypton calorimeter (LKr) and,
at large angles, by the lead-glass large-angle-veto (LAV) calorimeters.

  First results of the analysis of the 2016 data set w.r.t. the decay
  $K^+ \rightarrow \pi^+ \nu \bar{\nu}$ are presented at this conference \cite{Giuseppe}.
After a commissioning phase, NA62 is taking quality data towards this measurement since 2016
and the current
run of NA62 continues until the end of 2018, that is the end of `Run 2'.
NA62 then pauses with the pause of the CERN accelerator
infrastructure.
Restart of the experiment is expected in 2021 for `Run3'.

To understand NA62's capability for Exotics searches, it is helpful
to once again consider experiment and beamline shown in Fig.~\ref{fig:na62}.

Besides magnets, the `achromat' near 23m in the Fig. \ref{fig:na62}
comprises two move-able, $\sim$ 1.6m long blocks with a set of 
holes
allowing passage of the narrow beam and allowing adjusting its intensity.
These blocks are also dubbed `Target Attenuator eXperimental areas': TAXes.

During data-taking in the configuration with the beryllium target in
place (i.e. when the Kaon beam reaches the NA62 decay volume), 
a sizable fraction ($\sim$ 40 \%) of the protons 
pass through the target without interaction.
Thus these
impinge on the front, copper-part of the TAXes. These
protons are de-facto `dumped' and can be the source of so-far
undiscovered, weakly-interacting particles.

For this reason, during standard data-taking a number of parasitic trigger-lines have
been implemented which might help to detect the presence of
new particles. For example, an `exotic' multi-track 
trigger has been employed during 2017 data-taking,
built to trigger on events that did not originate from a Kaon decay.
This  exotic trigger
runs in parallel with a number of triggers optimized for
Kaon decays, notably $K^+ \rightarrow \pi^+ \nu \bar{\nu}$.

To foster the production of weakly interacting, novel
particles from dumped protons, NA62 can be run in `pure'
beam dump-mode by `closing' the upstream collimators and removing the
beryllium target.
`Closing the collimators/TAXes' means that
these are moved into a position
that completely blocks the primary SPS proton beam.
Followingly, in 2016 and continuing in 2017, the experiment 
has started to take data samples in this `pure' dump-mode
to
assess its capability and do first analyses for Dark Sector particles.
The statistics here is on the order of $\mathcal{O}(10^{16})$ POT.

In summary, NA62 can search for hypothetical, very weakly-interacting particles,
and thus illuminate the Dark Sector in at least three ways:
\begin{enumerate}
 \item Meson decays: New Physics Particles
 can be produced in decays of the Kaon (see examples in sect. \ref{sec:k_decay})
 \item Parasitic dump production:
 In `$\pi^+ \nu \bar{\nu}$' data taking, exotic particles  can be produced 
 by proton interactions
 far upstream the decay volume.
 This might be by  direct interactions of the primary particle
 in a target material or in the decay of secondaries.
 Weakly interacting new-physics particles can travel
 interaction-less up to the decay volume. If they decay
 away from the main beam-line they can be found and recorded by 
 the use of dedicated trigger lines, running parasitically to the trigger line
 for the main Kaon analysis (see examples in sect. \ref{sec:dump})
 \item Dedicated dump runs: To suppress backgrounds, NA62 can be run as `proton dump
 experiment', making it even more sensitive any particle of appropriate life-time
 potentially produced in upstream proton interactions. (see examples in sect. \ref{sec:dump})
\end{enumerate}

\section{NA62 results for exotics from Kaon decays \label{sec:k_decay}}

Let us first review some results and prospects of searches for exotic particles
in NA62 from Kaon decays.

\subsection{Invisibly decaying Dark Photons}

The decay chain $K^+ \rightarrow \pi^+ + \pi^0$, with $\pi^0 \rightarrow \gamma + A'$
has been investigated using 5\% of 2016 data.
This corresponds to about $1.5 \times 10^{10}$ Kaon decays.
Here, $A'$ is a Dark Photon (DP)  decaying invisibly (see sect. \ref{sec:dpdump} for visible DP decays).
The squared missing mass $m^2 = \left( p_K - p_{\pi} - p_\gamma \right)^2$,
peaks at 0 for the SM process $\pi^0 \rightarrow \gamma \gamma$
(where one of the $\gamma$s is lost).
By contrast, it should exhibit a peak around the $A'$ mass for the  $\pi^0 \rightarrow \gamma + A'$
decay, if the $A'$ is sufficiently strongly coupled given the statistics. 
A data-driven
background estimate, based on the tail with negative missing mass values, was used.
No statistically
significant excess has been observed and upper limits have been computed on the number
of signal events.

\begin{figure}[]
\begin{center}
 \includegraphics[width=0.6\textwidth]{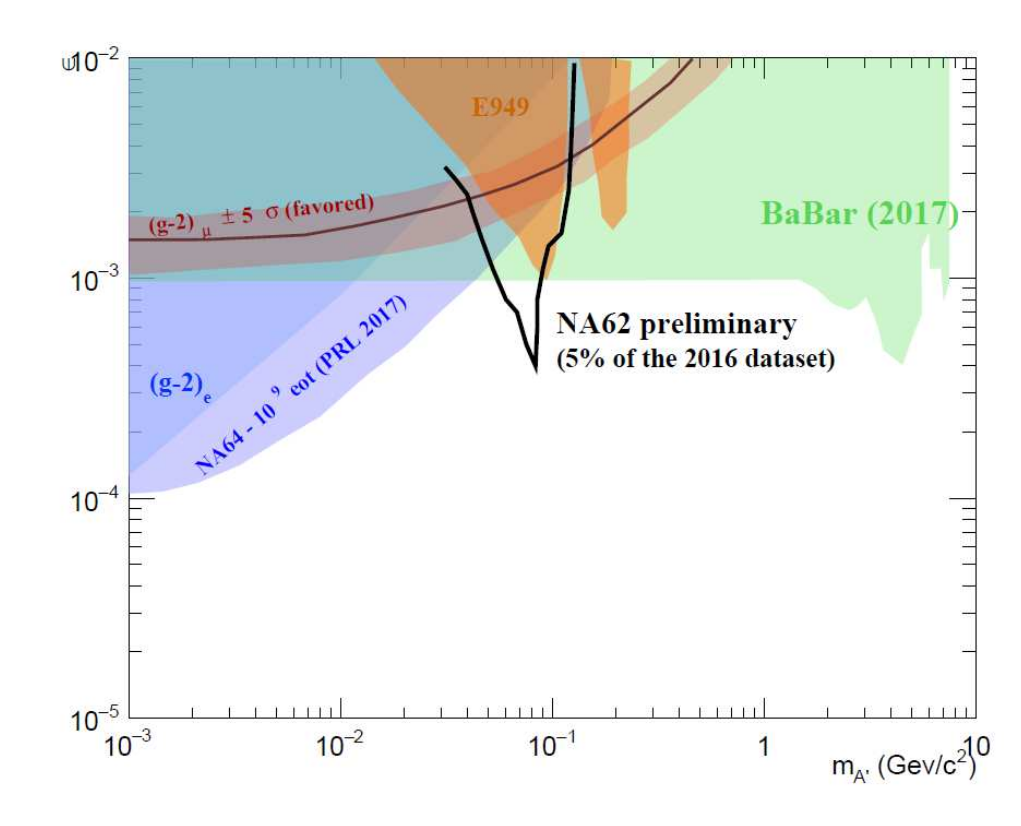}
 \caption{\label{fig:dpinv} \it
90\% CL exclusion limit in the coupling-mass plane for $\pi^0 \rightarrow \gamma A'$
with a Dark Photon $A'$ that decays 
into invisible final states.
The red band shows a  region of the parameters suggested to
explain the muon (g-2) anomaly (red band).
}
 \end{center}
 
 \end{figure}

The corresponding 90 \% confidence level exclusion limit on the kinetic mixing parameter 
versus the mass of the DP is shown in Fig. \ref{fig:dpinv} together with the limits from
BaBar, NA64 and E949, as compiled in \cite{Banerjee:2016tad}.

The ``stalactite-like'' shape shape of the NA62 exclusion region
 in fig. \ref{fig:dpinv} can be understood as follows:
Going to low $A'$ masses, the search is limited by the SM background
of $\pi^0$ decays with one photon lost.
On the other hand, at high masses, the limiting factor is kinematics.

\subsection{HNL from production search}

\begin{figure}[]
\begin{center}
 \includegraphics[width=0.55\textwidth]{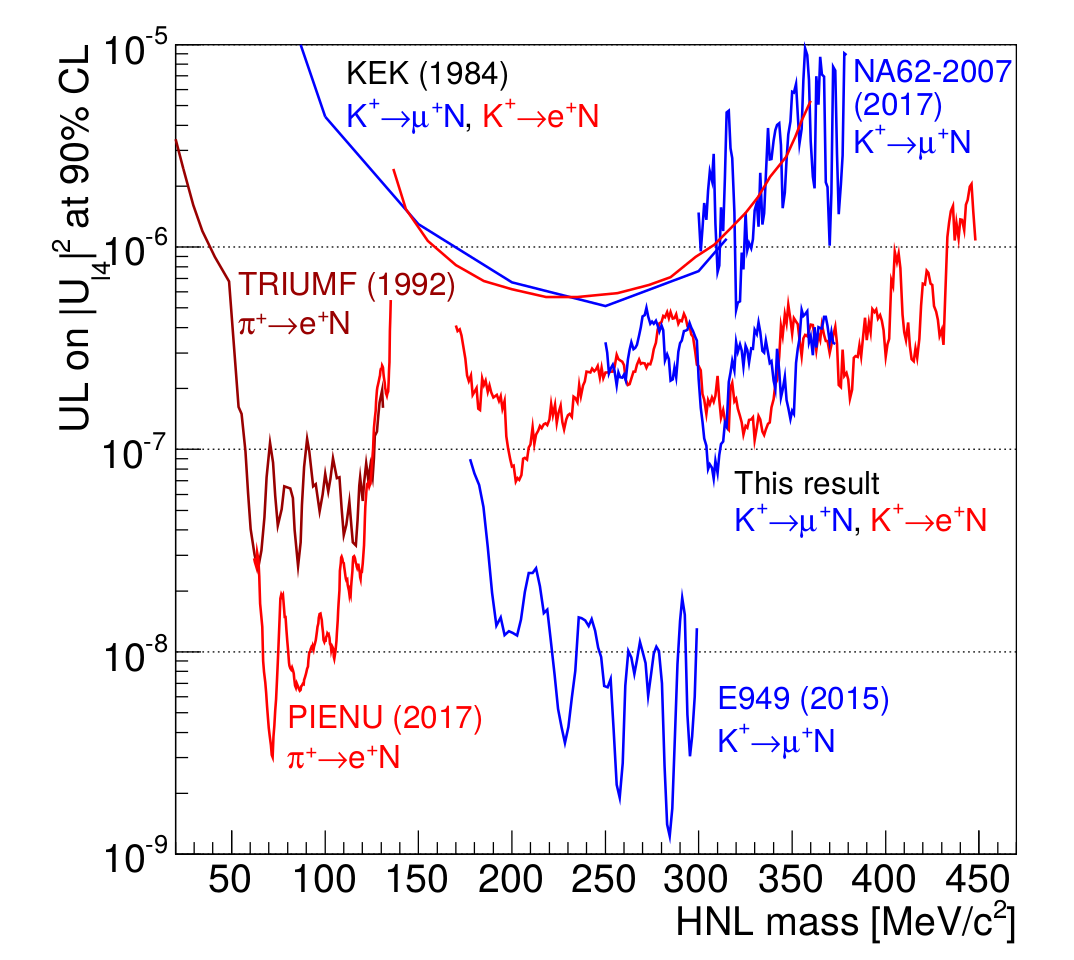}
 \caption{\label{fig:hnl} \it
90\% CL exclusion limit in the plane of coupling ($|U_{l4}|^2$) versus mass from various
pion and kaon leptonic decays. Limits from other production searches are 
shown together with the results from NA62 data taken in 2007 and 2015.
The latter is labeled
as ``This result'' \cite{CortinaGil:2017mqf}.
}
 \end{center}
 
 \end{figure}

A search for heavy neutral leptons (HNLs) 
that escape detection (see also section \ref{sec:hnl} for visible decay searches of HNLs) 
has been performed utilizing the decays
$K^{+} \rightarrow \mu^+/e^+ + N_{\mu/e}$, with $N$ the HNL. The analysis
proceeds through a ``bump-hunt'' in the 
positive squared missing mass region.
This analysis was based on
minimum-bias-triggered data from 2015 equivalent to $\sim 3 \times  10^8 K^+$ decays.
No signal has been observed and upper limits
have been placed \cite{CortinaGil:2017mqf}. This search considerably
improves the sensitivity with respect to previous experiments for both
the electron and muon modes above HNL masses $\gtrsim $300MeV, cf. Fig. \ref{fig:hnl}.
Note that the upper blue curve labeled NA62-2007 shows the limit from
\cite{Lazzeroni:2017fza}, based on data from 2007 (with the apparatus of the NA48/2 experiment) corresponding
to $\sim  10^7 K^+$  decays exploited to search for the leptonic muon mode.

\subsection{Further avenues for new particle searches from kaon decays}

The above-mentioned examples do not constitute
a comprehensive list of interesting channels
and the around $3 \times 10^{12} K^+$ collected in 2017 and the ongoing 2018 data taking
can be harvested for a plethora of new-physics signatures, see e.g. \cite{Bakhti:2017jhm,Fabbrichesi:2017vma}.
For example, as by-product of the main analysis, a very intriguing
possibility concerns flavored, ultralight axions such as the model of \cite{Calibbi:2016hwq},
or a bump hunt in $K^+ \rightarrow \pi^+ + X$, with $X$ decaying further to $l^- l^+$ with
$l= \mu, e$, see prospects in \cite{cds}.

\section{Prospects for upstream-produced exotics at $10^{18}$ POT at NA62 \label{sec:dump}}

In this section we discuss the prospects to
search for novel weakly interacting particles
produced upstream the NA62 fiducial volume
(TAX/target region in Figure \ref{fig:na62}).

For definiteness, we show the sensitivity
prospects for NA62 at $10^{18}$ POT for some new physics
models that are also considered for a much wider set 
of experiments in the context of CERN's `Physics
beyond collider' (PBC) studies \cite{PBChomepage}.
Thus, we heavily follow the benchmark sets provided in
\cite{pbcMaxim}.
All plots show a potential 90\% CL exclusion limit achievable by NA62,
if full background rejection can be achieved.

\subsection{ALPs with predominant photon coupling}

For axion-like particles (ALPs) as portal particles, one has
the possibility to write down couplings to gluons, quarks, leptons and other SM fields.
Here only the prospect of a strictly predominant coupling to
photons is shown (see \cite{Hochberg:2018rjs,Dolan:2014ska}
for some other possibilities).
If this is realized, the pre-dominant production
mechanism for ALPs would be via
Primakov production through photons-from-protons 
in the target (favored by a coherent $Z^2$ enhancement
over, e.g. an ALP-strahlung process \cite{Dobrich:2015jyk}).

The relevant interaction term including the ALP $a$ is
\begin{equation}
\mathcal{L}_{a, \rm int} = -\frac{1}{4} \, \ga \, \alp \, F^{\mu\nu}\tilde{F}_{\mu\nu} \; ,
\end{equation}  	
where $\ga$ denotes the photon ALP coupling.

\begin{figure}[]
\begin{center}
 \includegraphics[width=0.6\textwidth]{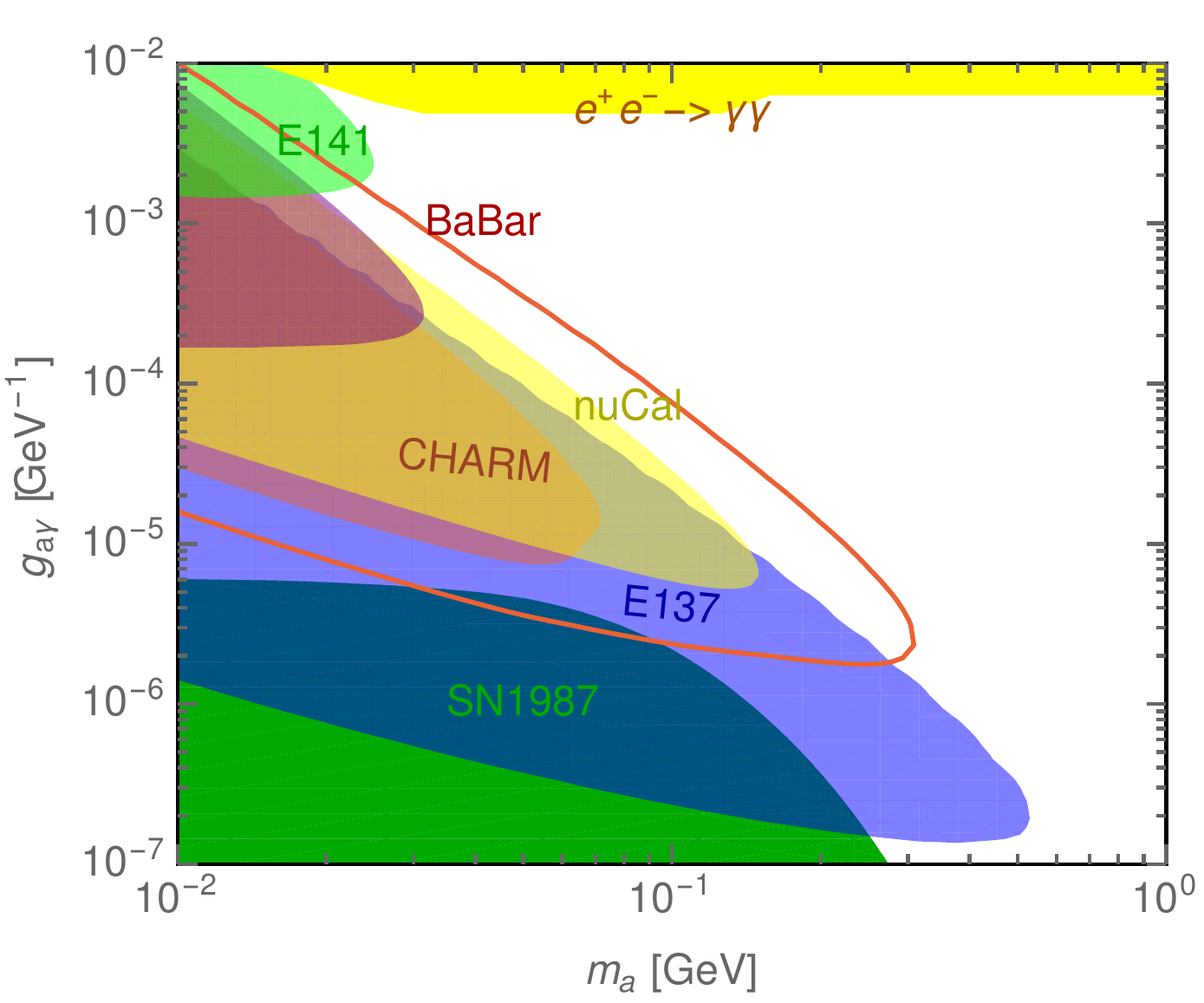}
 \caption{\label{fig:ALPs_status} \it
 Status of exclusions for ALPs coupled to photons in the MeV-GeV range.
The shaded areas correspond to excluded regions, the red line denotes the NA62 prospect at
$10^{18}$ POT.
}
 \end{center}
 
 \end{figure}

 The red, non-filled curve in
 Figure \ref{fig:ALPs_status} shows the prospects of a search performed
 at NA62 at $10^{18}$ POT.
 Other curves are taken from  \cite{Dobrich:2015jyk}  (with updates
 provided in \cite{Dolan:2017osp,Dobrich:2017gcm}).
 Projections are based on Primakov production
 through the equivalent photon approximation.
 In addition, detection of both photons from the ALP
 decay at a mutual distance of at least 10cm in the inner region of the LKr
 has been assumed in a toy MC.
 This toy MC has been cross-checked against the full NA62 MC.
 Note that this search needs to  strictly be performed in beam-dump-mode
 as no tracking for the photons is available.

\subsection{Higgs Portal}

\begin{figure}[]
\begin{center}
 \includegraphics[width=0.6\textwidth]{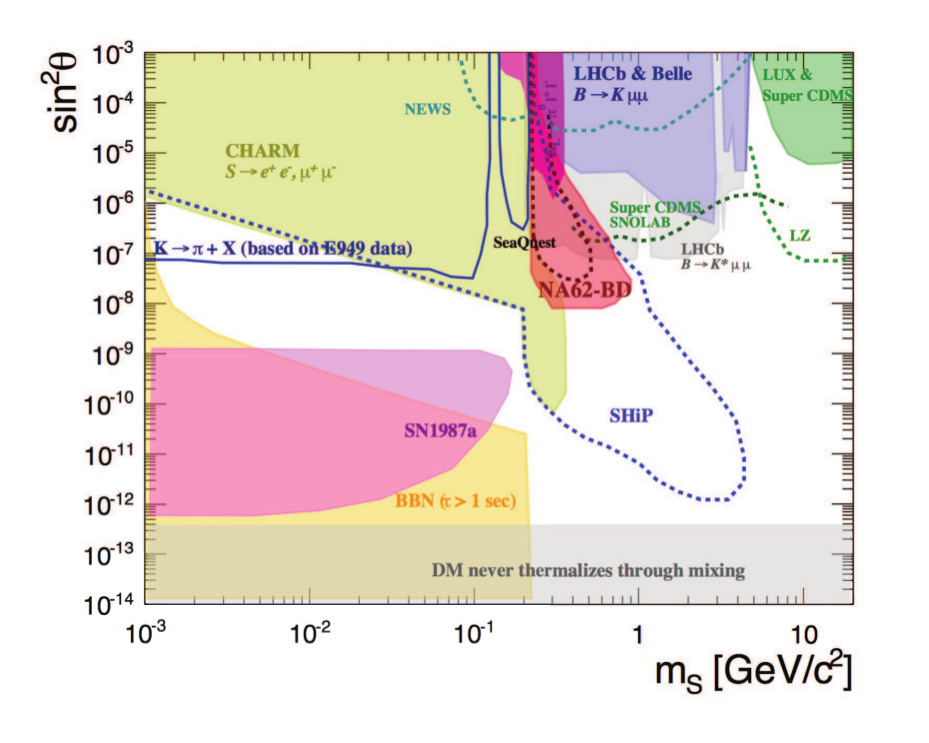}
 \caption{\label{fig:scalar_status} \it
 Status of exclusions for scalars mixing with the Higgs
 as described in the text. Plot as in \cite{Krnjaic:2015mbs}
 with NA62 sensitivity projection in red, labeled `NA62-BD'.
 Also indicated are projections for SeaQuest and SHiP.
}
 \end{center}
 
 \end{figure}

In the following we project the sensitivity for 
NA62 for a scalar $S$ with an interaction

\begin{equation}
\mathcal{L}_{\rm int, scalar}\sim \mu S  H^\dagger H \ \  , \theta  =  \mu v/(m_h^2-m_S^2) \ ,
\end{equation}  
where $v$ is the Higgs VEV and a mixing parameter $\theta$, valid 
for small mixings \cite{Krnjaic:2015mbs} is introduced.
Above we have omitted the possibility of an additional $S^2$ interaction.
Our projection for NA62 at $10^{18}$ POT is given in Fig. \ref{fig:scalar_status}.

The dominant production mode here is from B-Mesons produced in the dump,
which subsequently decay into final states with $S$ particles.
The decay of $S$  has been evaluated in a toy MC cross-checked against the full MC
and considers final states $\mu \mu$, $e e$, $\pi \pi$, $K K$.

\subsection{Dark Photons \label{sec:dpdump}}

The Dark Photon (vector) portal considered here is a minimal model in which an additional U(1)
is introduced that mixes with the SM photon:
\begin{equation}
\mathcal{L}_{\rm int}\sim \frac{\epsilon}{2 \cos(\theta_W)} F'^{\mu\nu}{B}_{\mu\nu} \; ,
\end{equation}  	
with $\epsilon$ being the kinetic mixing.

\begin{figure}[]
\begin{center}
 \includegraphics[width=0.6\textwidth]{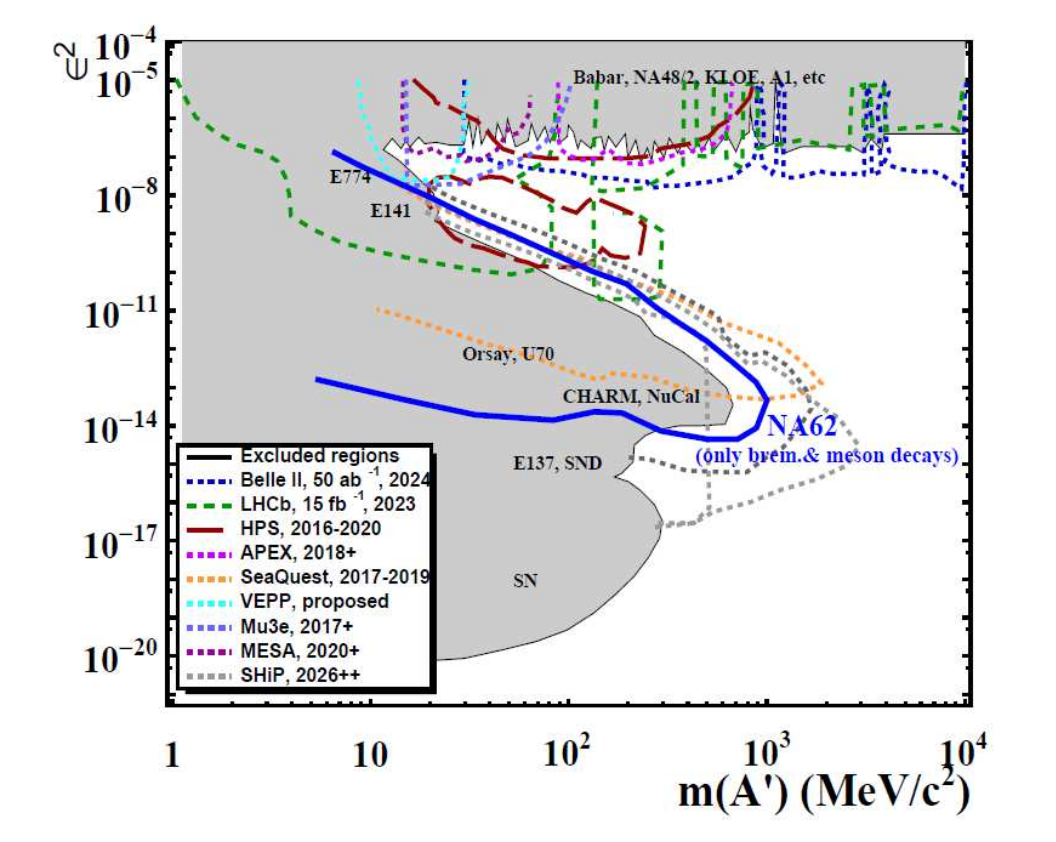}
 \caption{\label{fig:DP_status} \it
 Status of exclusions for Dark Photons.
 The blue line shows the sensitivity projection of NA62
 based only on production of Dark Photons from Meson decays and Bremsstrahlung at $10^{18}$ POT
 from the Beryllium target.
}
 \end{center}
 
 \end{figure}

Figure \ref{fig:DP_status} shows the current state
of exclusions together with the prospect sensitivity of NA62.
Final states in $ee$, $\mu \mu$ have been considered, 
and plausible trigger and selection efficiencies have been accounted for.
A toy MC has been set up and cross-checked against the full MC.

For this projection, only Dark Photon production
via Meson decays of $D's, \pi^0, \eta, \eta', \Phi, \rho, \omega$
and in Bremsstrahlungs-production has been considered.
Additionally this production is assumed to take place in the Beryllium target only.
%Also this projection takes into account trigger efficiencies.

Considering production in the more downstream TAX will improve the
projection further. Also, in principle, QCD processes such
as $q \bar{q} \rightarrow A'$ can contribute, especially at higher masses \cite{deNiverville:2012ij}.
Albeit this is plagued by theoretical uncertainties in the corresponding
production cross-section. Such processes are not considered here.
In this sense, our projection is rather  conservative.

\subsection{Heavy Neutral Leptons \label{sec:hnl}}

For the neutrino portal 
\begin{equation}
\mathcal{L}_{\rm int, HNL}\sim \Sigma F_{\alpha i} ( \bar{L}_\alpha H) N_i
\end{equation}  
with the sum over HNLs, $N$ and the flavor of lepton doublets $L$. $F$ denotes Yukawa couplings.
More details can be found in \cite{Gorbunov:2007ak,Alekhin:2015byh}.

\begin{figure}[]
\begin{center}
 \includegraphics[width=0.99\textwidth]{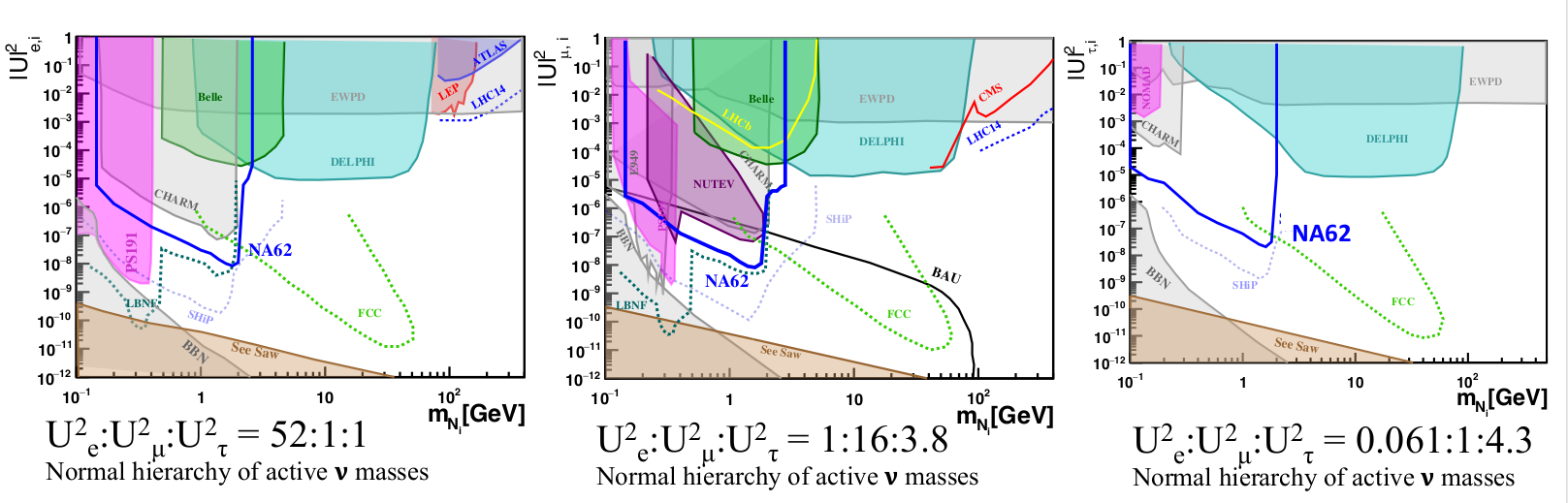}
 \caption{\label{fig:HNLs} \it
 Status of exclusions for HNLs coupled in the MeV-GeV range (together with some projections of 
 a selected set of other experiments).
 In the plots, three extreme coupling scenarios are addressed separately,
 cf. \cite{Gorbunov:2007ak,Drewes:2018gkc}. The blue lines show the prospect
 of NA62 to test these scenarios if full background rejection can be achieved.
}
 \end{center}
 
 \end{figure}

Figure \ref{fig:HNLs}
shows NA62 prospect sensitivity \cite{Drewes:2018gkc}
 for $10^{18}$ POT in the coupling
versus mass plane for a three theoretical scenarios of heavy neutral lepton models, corresponding 
to the highest possible couplings to electrons (left-most panel), muons (central panel), and
taus (right-most panel) and normal neutrino hierarchy.

\section{A word on background rejection}

\begin{figure}[htb]
    \begin{center}
        {\includegraphics[scale=0.29]{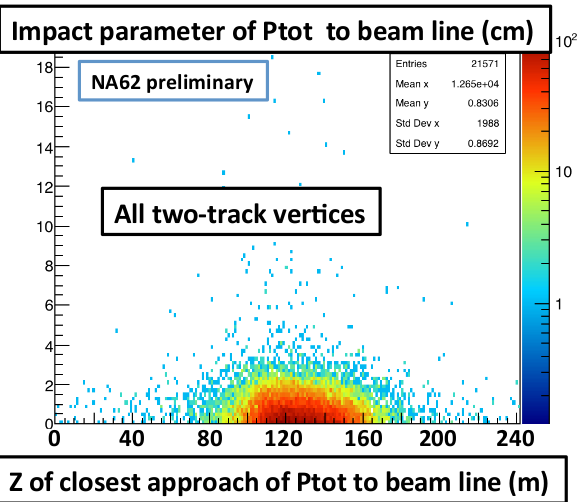}}\hspace{0.5cm}
      {\includegraphics[scale=0.265]{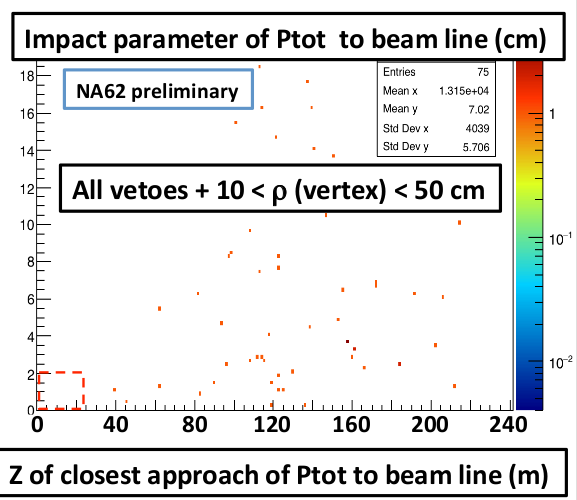}}
        \caption{\it Left-hand Plot: data corresponding to $10^{15}$POT,
        after quality cuts described in text. Shown is the distance of the extrapolated
        total vertex momentum to the nominal beam-line  at the position
        of the closest approach to the beam-line. Right-hand plot:
        Result after additional veto conditions and the requirement that
        the distance of the vertex to the nominal beamline is between 10 and 50cm.
        The red-dashed `signal box' has no remaining event.}
\label{fig:bkg}
    \end{center}
\end{figure}

The strategy and performance of background rejection for charged
final states (Dark Photons, Scalars, and HNLs as discussed above) can be to some extent
understood from fig. \ref{fig:bkg}. In it, we show some results obtained in
in parasitic mode  (see also \cite{tomPBC}). This sample is taken
is during nominal data-taking, in a parasitic
trigger stream that requires two coincident muons (10ns) window
as well as an energy in the LKr calorimeter $<20$ GeV.
The statistics is $\sim 10^{15}$ POT.

On the l.h.s. of fig. \ref{fig:bkg} we show the distance of
the extrapolation of the total vertex momentum to the nominal beam-line at the closest
approach to the nominal beam-line. All two-track vertices are
 shown after 
a number of quality and track acceptance cuts.
As expected, the majority of vertices and thus the background for exotic final states 
comes from $K$ and $\pi$ decays in and before the fiducial region.
One can employ however, that the number of such vertices
decreases steeply as one moves away from the beam-line.

The right-hand side of Figure \ref{fig:bkg} shows the same data after 
a number of additional veto conditions, including the requirement
that the vertex is located in the fiducial volume
in between a $z$ of 105m and 165m: Most importantly,
the r.h.s. requires that the 2-d vertex distance $\rho$ 
is  between 10 and 50 cm from the beamline.

As can be seen, for this data set, no event is compatible with stemming
from an exotic particle produced in the region between the
NA62 Beryllium target and the TAX collimators in between 0 and $\sim25$m:
The extrapolation of the total vertex momentum in the red-dashed signal
box contains no entry.

Note, that for `pure dump runs', a similar reasoning/analysis can be applied.

For the situation of fully neutral final states (such as ALPs) the above analysis is not useful,
but in such a situation, it is feasible
to exploit the fact that ALPs that exist in a still un-explored
parameter region are necessarily very boosted when they reach the NA62 sensitive volume \cite{tomPBC}.
The probability to reach decay volume for an ALP is  $\sim \exp(-l_{\rm absorber}/l_{d})$, where
$l_{d}= \gamma \beta \tau \sim \frac{E_a}{m} \frac{64 \pi}{m^3 g^2}$
and the ``absorber'' length for NA62 is $l_{\rm absorber} \simeq 81$m.
Following fig. \ref{fig:ALPs_status}, the yet-to-be discovered ALPs of interest 
have a short live-time (comparably large couplings) and will arrive at the NA62 sensitive volume at high $E_a$.

\section{Conclusion: NA62 now and in run 3}

The NA62 experiment, aimed at the measurement
of $K^+\rightarrow \pi^+ \nu \bar{\nu}$ has
released its first analysis on this decay channel based on 2016 data,
validating its strict performance requirements on the detector.
2017 data is being analyzed and the 2018 run ongoing.
To reach a measurement of $K^+\rightarrow \pi^+ \nu \bar{\nu}$ 
with a satisfactory precision, NA62 aims to continue to take data
after LS2.

With this data, also exotic searches from Kaon decays can be performed. 
Recent results on production searches for heavy neutral
leptons and preliminary results on invisible decays of dark photons were shown in this article.

The requirement of hermetic coverage and
$\mathcal{O}(100)$-ps timing resolution allows also for a number
of searches for new particles, potentially residing in a `Dark Sector' being
produced in the upstream TAX collimator.
In these proceedings we have shown the prospects for some of these models (see
e.g. \cite{Renner:2018fhh} for other possibilities)
at $\sim 10^{18}$ POT. This corresponds to a $\sim$ 1-year long data taking.

For this reason, the
NA62 collaboration is currently discussing the possibility to use a fraction of the beam
time during Run 3 (2021-2023) to operate NA62 with closed upstream collimators (beam-dump).
The current NA62 run is
exploited to evaluate background rejection capability and
perform first searches for new physics for some of the presented channels.

\section{Acknowledgements}
The author would like to thank the organizers of the
``Vulcano Workshop 2018 - Frontier Objects in Astrophysics and Particle Physics''
for a thematically wide and  intriguing conference in a wonderful setting.

\end{document}